\documentclass[traditabstract,letter]{aa}
\usepackage{graphicx}                    
\usepackage{bm}
\usepackage{times}                    
\usepackage{color}                       
\usepackage{url}                         

\def\beg{\begin{eqnarray}}
\def\ende{\end{eqnarray}}
\def\lsim{\lower.4ex\hbox{$\;\buildrel <\over{\scriptstyle\sim}\;$}}

\renewcommand{\vec}[1]{\mbox{\boldmath $#1$}}
 
\def\Om{{\it \Omega}}
\def\R{{R\"udiger}}

\begin{document}



\title{The existence of the $\Lambda$ effect in the solar convection zone  indicated by SDO observations}
\titlerunning{Observational evidence for the existence of the $\Lambda$ effect}
%
\author{G.~R\"udiger \and M.\ K\"uker \and I. Tereshin}


%
  \institute{Leibniz-Institut f\"ur Astrophysik Potsdam, An der Sternwarte 16, D-14482 Potsdam, Germany,
                     email:  gruediger@aip.de, mkueker@aip.de}

\date{Received; accepted}
 
\abstract{The empirical  finding with data from the Solar Dynamics Observatory (SDO)  of positive  (negative)  horizontal Reynolds stress at the northern (southern) hemisphere for solar giant cells (Hathaway  et al. 2013) is discussed for its consequences for the 
theory of the solar/stellar differential rotation. Solving the nonlinear  Reynolds equation for the angular velocity under neglect of the meridional circulation we show that 
the horizontal Reynolds stress of the northern hemisphere  is always negative at the surface  but it is positive in the bulk of the solar convection zone by the action of the $\Lambda$ effect. The $\Lambda$ effect, which describes the angular momentum transport of rigidly rotating anisotropic turbulence and which avoids a rigid-body rotation of  the  convection zones, is in horizontal direction  of cubic power  in $\Om$ and it is always  equatorward directed. Theories  without $\Lambda$ effect which may also  provide the observed solar rotation law only by the action of a  meridional 
 circulation  lead to  a  horizontal Reynolds stress with the opposite sign as observed.
}


%
\keywords{Sun: convection -- Sun: rotation -- giant cells  }

\maketitle
%
\section{Introduction} \label{Section1}
The nonrigid rotation of the stellar surfaces which currently has been observed for many thousands of stars (Reinhold et al. 2013) 
demonstrates the existence of an effective angular momentum transport in rotating convection zones. There are  various transporters of angular momentum in a rotating turbulence field. The turbulence-induced Reynolds stress transports  by its components $\langle u_r u_\phi  \rangle$ and  $\langle u_\theta u_\phi  \rangle$  angular momentum in radial  and in latitudinal directions. Boussinesq (1897) and  Taylor (1915)   connected the one-point  correlation tensor $Q_{ij}=\langle u_i(\vec{x},t)u_j(\vec{x},t) \rangle$, which is symmetric by definition in its indices $i$ and $j$, with the shear of a large-scale  flow $\vec{U}$ so that 
$
Q_{ij}= ...-\nu_{\rm T}(U_{i,j}+U_{j,i})
$
results with the positive-definite eddy viscosity $\nu_{\rm T}$.  
Here the notation  $\vec{U} +  \vec{u}$  as  the fluctuating
 velocity  field with the  background flow   $\vec{U}$ has been used.  This ansatz which does not reflect the  anisotropies in the turbulence field immediately yields
\begin{equation}
Q_{r\phi}= -\nu_{\rm T} r \sin\theta\frac{\partial \Om}{\partial r}, \ \ \ \ \ \ \ \ \ \ \ \ Q_{\theta\phi}= -\nu_{\rm T}  \sin\theta\frac{\partial \Om}{\partial \theta}
\label{Qphi}
\end{equation}
for the  mentioned cross-correlations.   As they vanish for rigid rotation they cannot serve to maintain nonuniform  rotation $\Om=\Om(r,\theta)$. The turbulence  in a convection zone, however, is subject to a distinct radial anisotropy by the central gravity which questions the validity of the  relations (\ref{Qphi}). Kippenhahn (1963)  with the concept of an anisotropic viscosity tensor derived a generalized linear relation for the radial transport, i.e. 
\begin{equation}
Q_{r\phi}= \nu_{\rm T}\left(- r \frac{\partial \Om}{\partial r} + V(r)\Om\right) \sin\theta,
\label{Qphi1}
\end{equation}
where the second term in (\ref{Qphi}), however,   remains unchanged\footnote{One finds more  details to the history of  `anisotropic viscosity'  and  
the related references  in \R\ (1989)}. 

Stationary solutions for the angular velocity $\Om$ have to fulfill the Reynolds equation
\beg
\frac{1}{r^2}\frac{\partial \rho r^3 Q_{r\phi}}{\partial r}+ \frac{1}{\sin^2\theta}\frac{\partial\rho\sin^2\theta Q_{\theta\phi}}{\partial\theta}=0,
\label{Rey}
\ende
which ensures the conservation of the angular momentum in the rotating turbulence under neglect of the mean meridional circulation and     
the molecular viscosity. Here $\rho$ is the  density. Inserting (\ref{Qphi})$_2$ and  (\ref{Qphi1}) into 
(\ref{Rey}) one  finds $\Om$ 
dependent on $r$ but independent of $\theta$.  For positive radial shear in the rotation law, on the other hand,   a meridional flow {\em towards the equator} is induced at the surface which    accelerates the equator by transporting angular momentum. Such a  meridional flow, however, has not been observed so far.  In contrast, the observed slow meridional flow  rises  at the equator and flows in polar direction  along the surface (Gizon \& Rempel 2008; Schad et al.  2012). 

\begin{figure}[h]
\begin{center}
\includegraphics[width=\columnwidth]{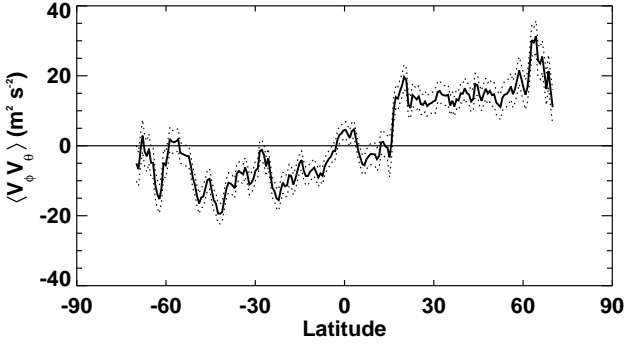}
\end{center}
\caption[]{The horizontal  Reynolds stress observed with HMI on the NASA  Solar Dynamics Observatory by Hathaway et. al. (2013). With   permission by D.H. Hathaway. }
\label{fig0}
\end{figure}

After  (\ref{Qphi})$_2$ 
an accelerated equator provides  a negative (positive)  horizontal Reynolds stress  at the northern (southern) hemisphere. Recently,  Hathaway et al. (2013) from the data of  the NASA Solar Dynamics Observatory (SDO) empirically isolated a giant cell pattern at the solar surface where the  proper motions form a horizontal
cross-correlation $Q_{\theta\phi}$ antisymmetric with respect to the equator and with a positive  amplitude of $2 \cdot 10^{5}$ cm$^2$/s$^2$ at northern mid-latitudes (Fig. \ref{fig0}). Cells with faster
 rotation tend to move equatorward and vice versa. Because of the symmetry conditions  for the horizontal cross correlation (vanishing at poles and equator by definition) it makes sense to introduce the dimensionless factor $W$ by means of 
\beg
Q_{\theta\phi}= \nu_{\rm T}\Om_0  \cos\theta \sin^2\theta W.
\label{Ward}
\ende
The product $ \nu_{\rm T}\Om_0$  ($\Om_0$ the characteristic angular velocity) is the scalar with the correct dimension which reflects the fact that only 
 rotating  turbulence possesses finite horizontal cross correlations $Q_{\theta\phi}$. After Hathaway et al. (2013) we have $W\simeq 0.3/\nu_{12}$ with $\nu_{12}=\nu_{\rm T}/(10^{12}{\rm cm}^2/{\rm s}^2)$. Previous investigations with magnetic tracers led to values for $W$ higher by two orders of magnitudes (Ward 1965; Gilman \& Howard 1984; Balthasar et al. 1986) while the statistical analysis of coronal bright points yielded  smaller numbers  (Vrsnak et al. 2004). Note that for solitary spots  Nesme-Ribes et al. (1993) 
 even find small but negative correlations.

The empirical results obviously indicate the invalidity of the simple Boussinesq relation (\ref{Qphi})$_2$ for rotating convection. The observed equatorial acceleration -- if it is a deep-seated phenomenon -- would always provide negative (positive) cross-correlation  values for the northern (southern) hemisphere. As this is not observed it
should be natural to replace the expression  (\ref{Qphi})$_2$ by
\beg
Q_{\theta\phi}= \nu_{\rm T}\left( - \sin\theta\frac{\partial \Om}{\partial \theta}+ \sin^2\theta\cos\theta H (\frac{\Om}{\Om_0})^2\Om\right),
\label{Qt}
\ende
where  the second term on the RHS  is of cubic order in $\Om$. The counterpart 
of (\ref{Qt}) with respect to the radial transport of angular momentum is 
\beg
Q_{r\phi}=&&\nonumber\\
 &\nu_{\rm T}\left(- r \sin\theta\frac{\partial \Om}{\partial r}+ \sin\theta V\Om-\sin\theta\cos^2\theta H (\frac{\Om}{\Om_0})^2\Om\right)&,
\label{Qr}
\ende
where  here  also a term $V$ linear in $\Om$ appears. All the nondiffusive terms in the expressions for the turbulent angular momentum transport 
can be written in the tensorial form $Q_{ij}=...+\Lambda_{ijk}\Om_k$ with the $\Lambda$ effect tensor $\Lambda_{ijk}$ symmetric in its first two indices   which is even in $\Om$ by definition. This is why all terms in (\ref{Qt}) and  (\ref{Qr}) are odd in $\Om$. A tensor of 3$^{\rm rd}$  rank  even in $\Om$ can only be constructed in turbulent fluids if the turbulence is anisotropic by itself and/or inhomogeneous as a consequence of the density stratification. Both conditions are fulfilled in stellar convection zones.

We have shown earlier by means of quasilinear turbulence theory that the function $V$ is negative and  exists mainly in the supergranulation layer while $H$  is positive and  exists mainly in the bulk of the convection zone. Numerical simulations provide very similar results (Chan 2001; K\"appyl\"a et al. 2011; for more details see \R\ et al. 2013). We also know that  the positive  quantity $H$ in (\ref{Qt}) and (\ref{Qr}) both  contribute to the equatorial acceleration so that it is by far not trivial to find the sign of (\ref{Ward}) as a function  of depth.

Solving the equation (\ref{Rey}) with the expressions (\ref{Qt}) and (\ref{Qr}) leads for a simple model with weak shear and uniform density,  uniform $V$ and $H$ and for stress-free  boundaries ($Q_{r\phi}=0$)  to the  the estimates
\beg
\frac{\delta \Om}{\Om_0}\simeq \frac{1+d}{2} \ H,  \ \ \ \ \ \ \ \ \ \ W\simeq - d H
\label{old}
\ende
  (thin shells) for the normalized equator-pole-difference of  $\Om$ and the amplitude of  $W$ both taken at the surface. Both observable  values do not depend on the viscosity value. Here $d\simeq 0.3$ is the normalized thickness of the convection zone.  The negative sign of $W$ means that the latitudinal shear at the surface (due to the action of  $H$) always exceeds the value of $H$.  However, the data of Hathaway et al. (2013) 
 concern the giant cell pattern which exists deep in the convection zone rather than at its surface. The shear there is reduced by the lower boundary condition (tachocline!) 
 so that  $W$ becomes   positive. The analytical relation for $W$   at the bottom of the convection zone reads $W\simeq d H>0$ because of the reduction of the latitudinal shear.   The observed positivity of the quantity $\cos\theta Q_{\theta\phi}$ appears to be compatible with the existence of the solar tachocline which rapidly reduces the equator-pole difference to zero (probably by  Maxwell stress).
 \section{Mean-field models}\label{Section2} 
\begin{figure}[hbt]
\begin{center}
\mbox{
\includegraphics[width=4.5cm,height=4.4cm]{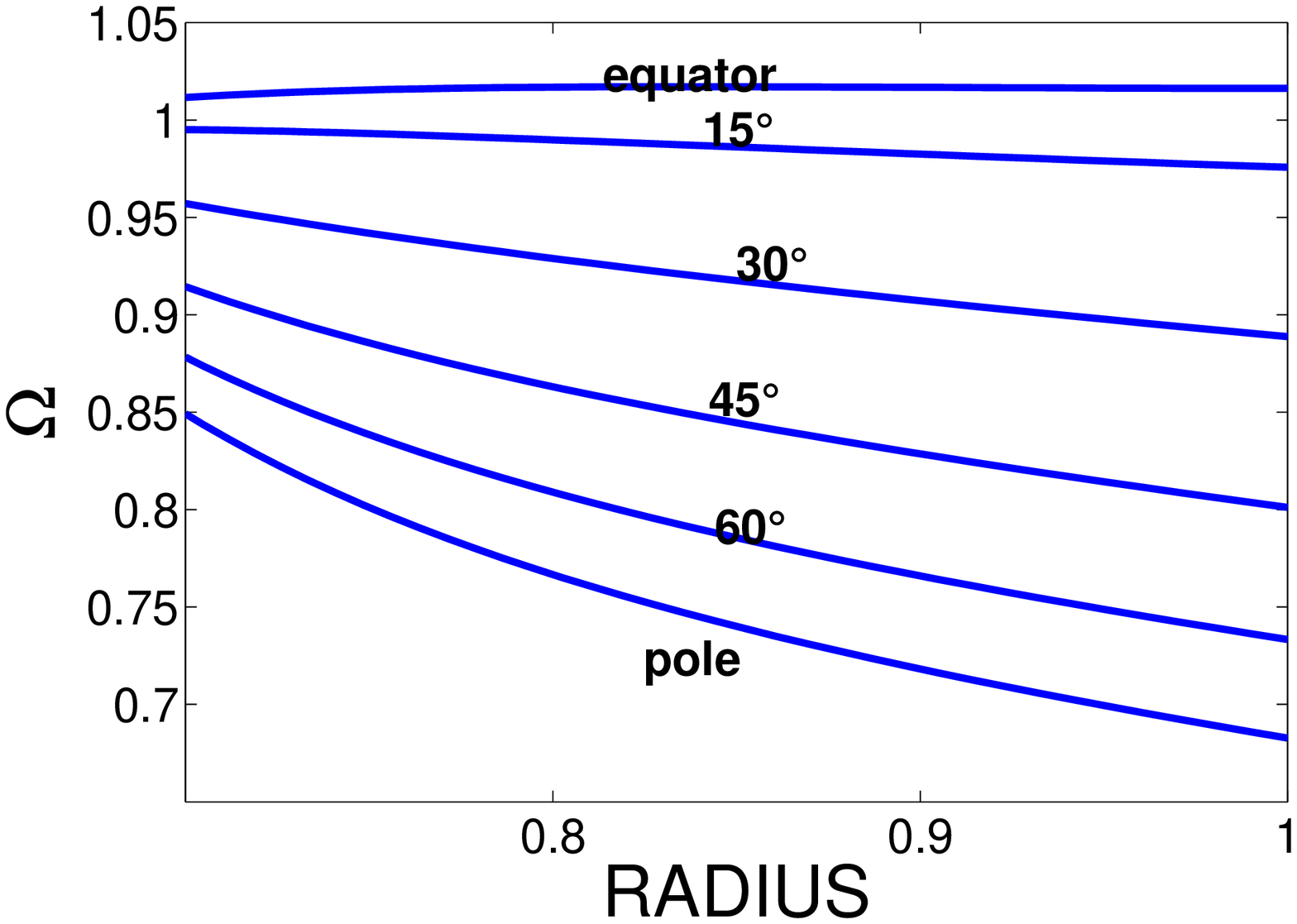}
\includegraphics[width=4.5cm,height=4.4cm]{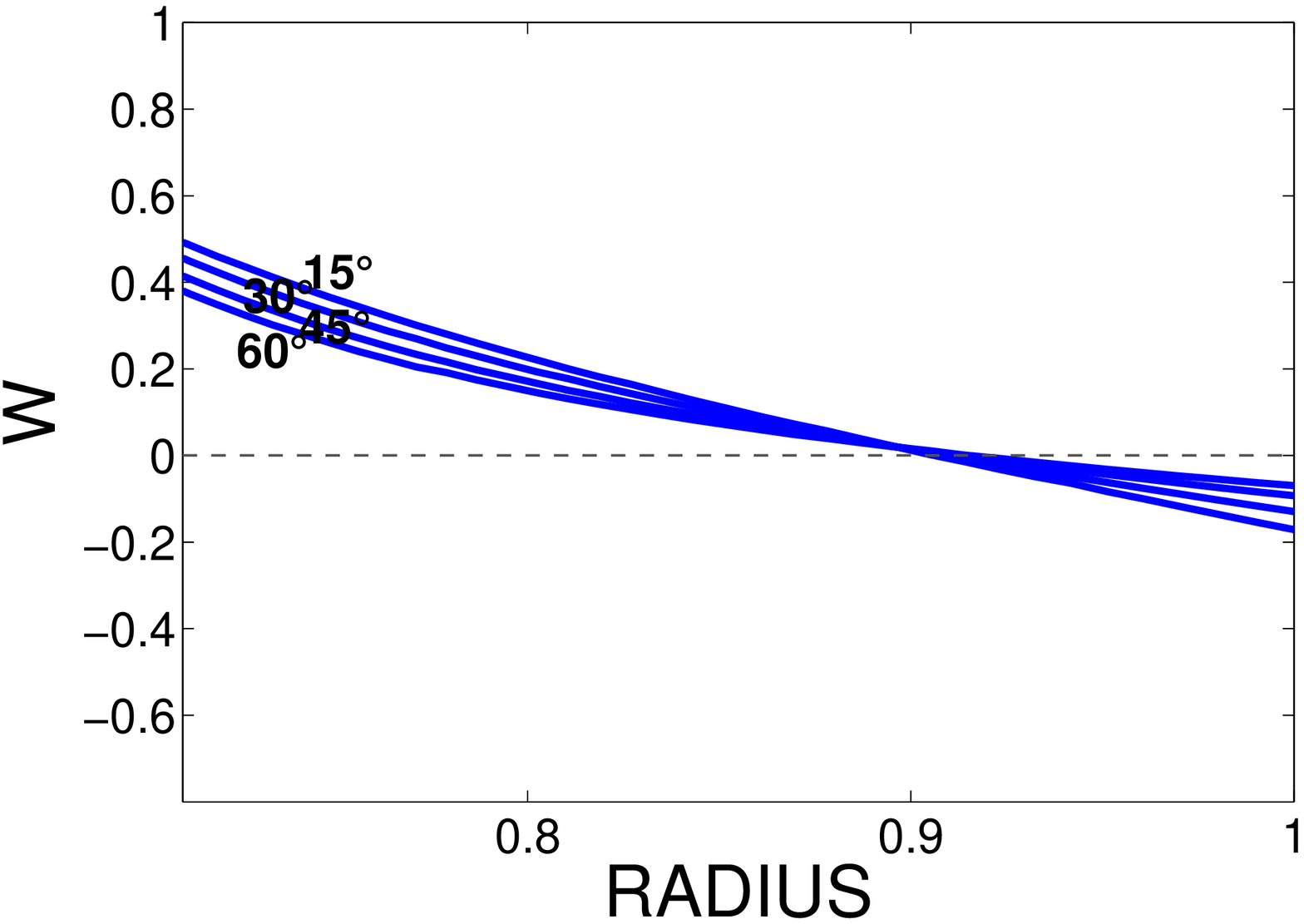}}
\end{center}
\caption[]{The rotation law (left) and the horizontal  Reynolds stress function $W$ (right) for a mean-field model with $V=0$ and  $H=1$  in the entire convection zone ($x_{\rm in}\leq x\leq 1$). Here $x$ is the fractional radius of the star. Density is assumed as uniform.}
\label{fig1}
\end{figure}

\begin{figure}[hbt]
\begin{center}
\mbox{
\includegraphics[width=4.5cm,height=4.4cm]{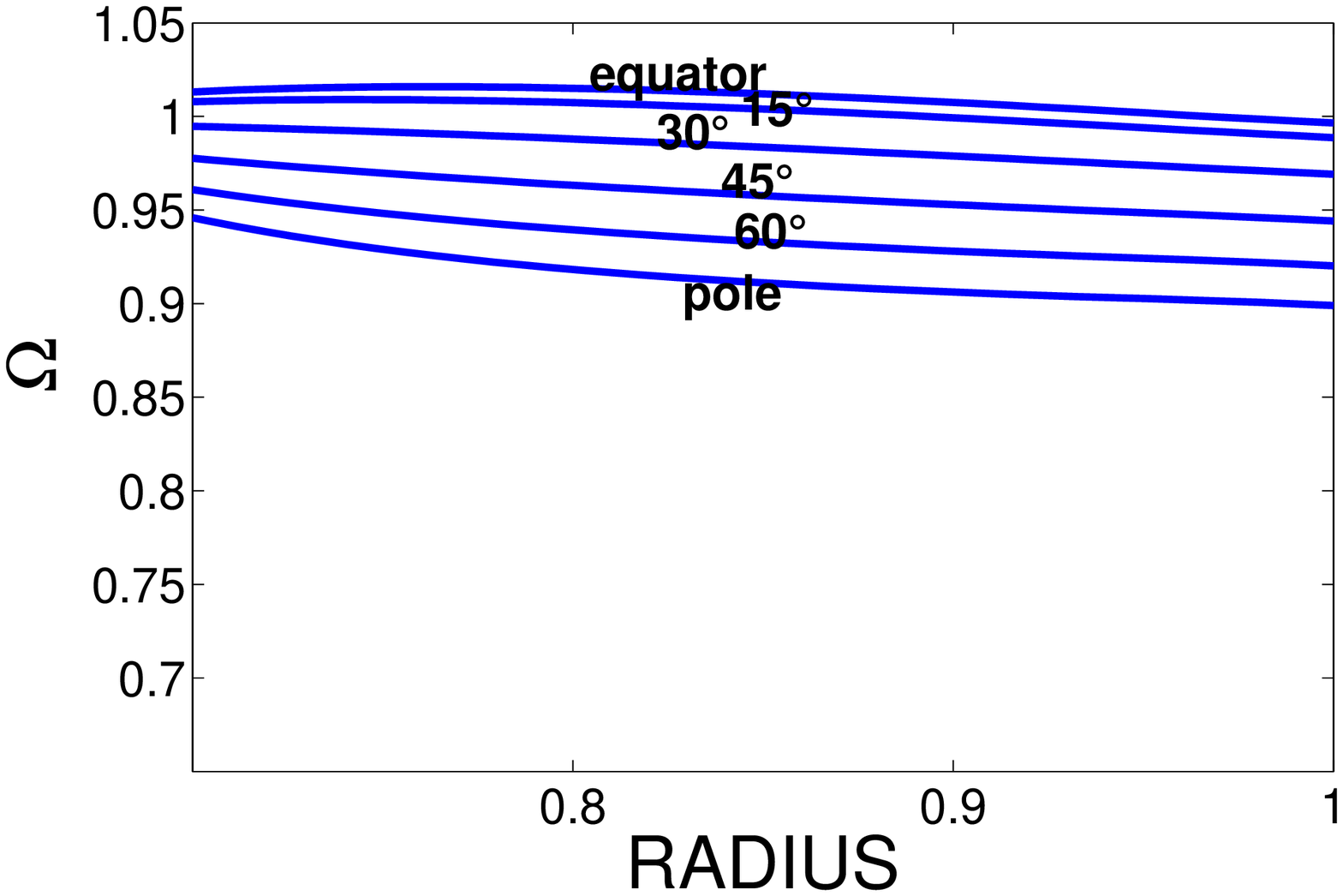}
\includegraphics[width=4.5cm,height=4.4cm]{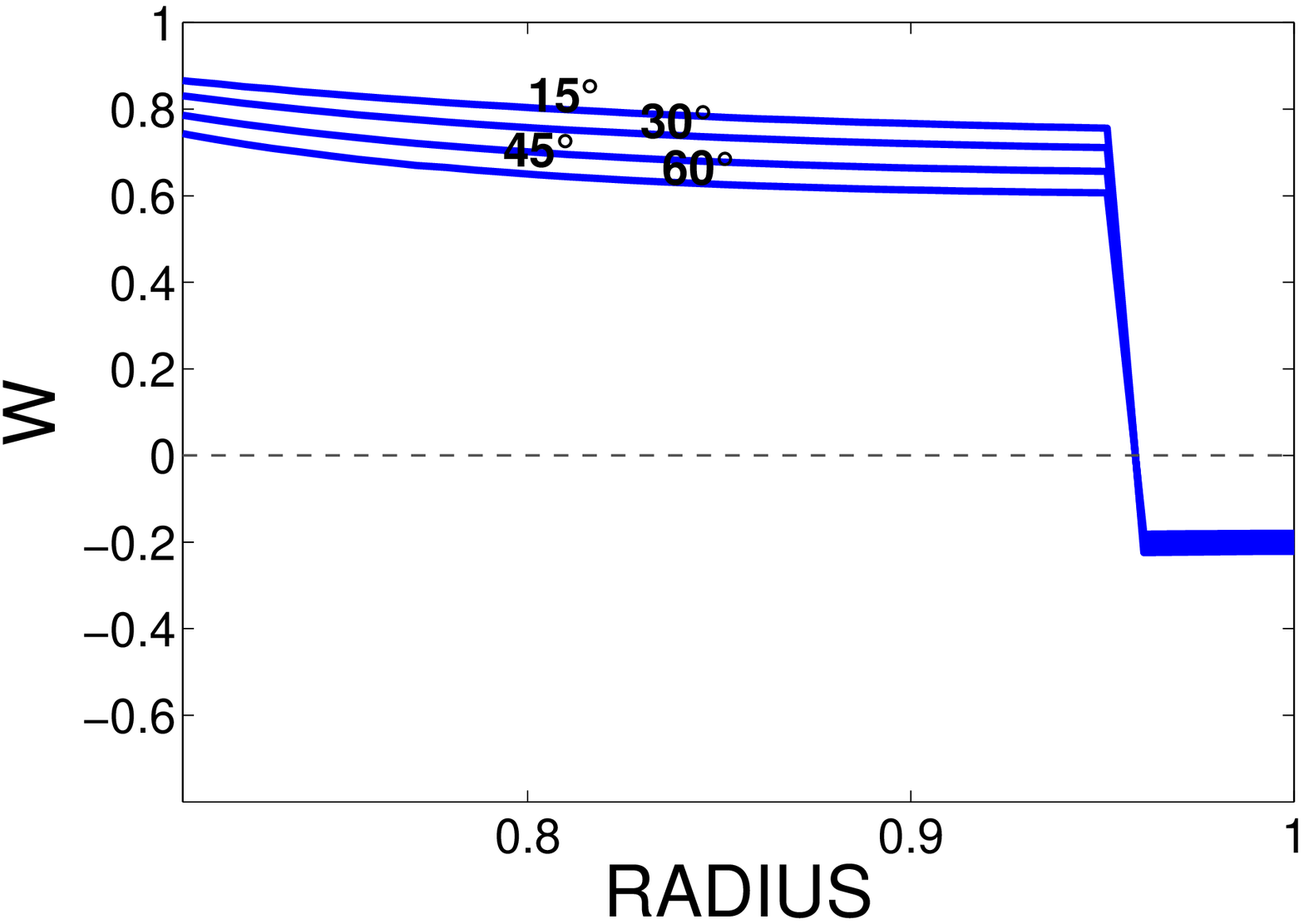}}
\end{center}
\caption[]{The same as in Fig. \ref{fig1} but  with $H=1$ only  in the bulk of the convection zone ($x_{\rm in}\leq x\leq 0.95$) and $V=-0.1$  for $x>0.95$.}
\label{fig2}
\end{figure}

\begin{figure}[hbt]
\begin{center}
\mbox{
\includegraphics[width=4.5cm,height=4.4cm]{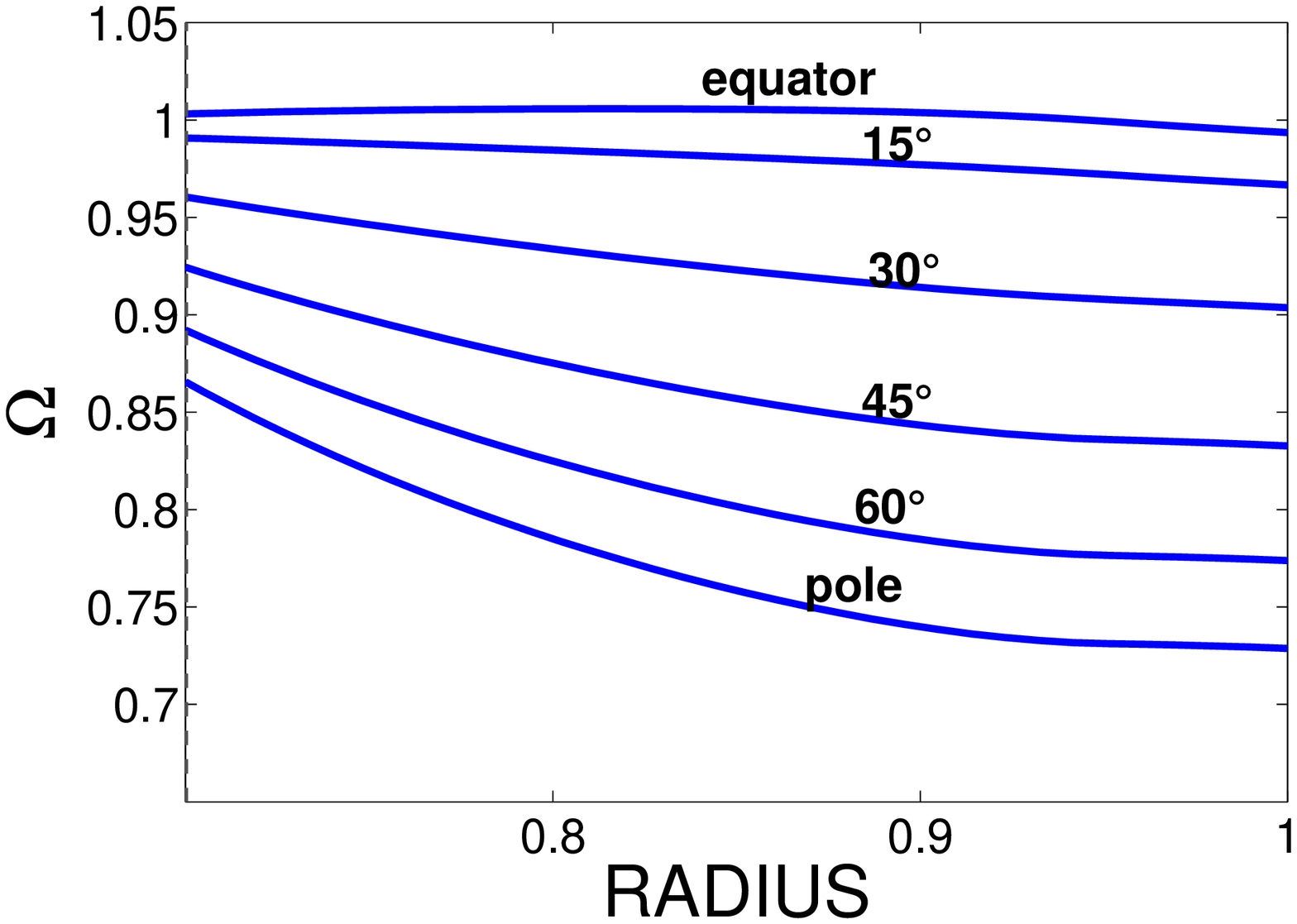}
\includegraphics[width=4.5cm,height=4.4cm]{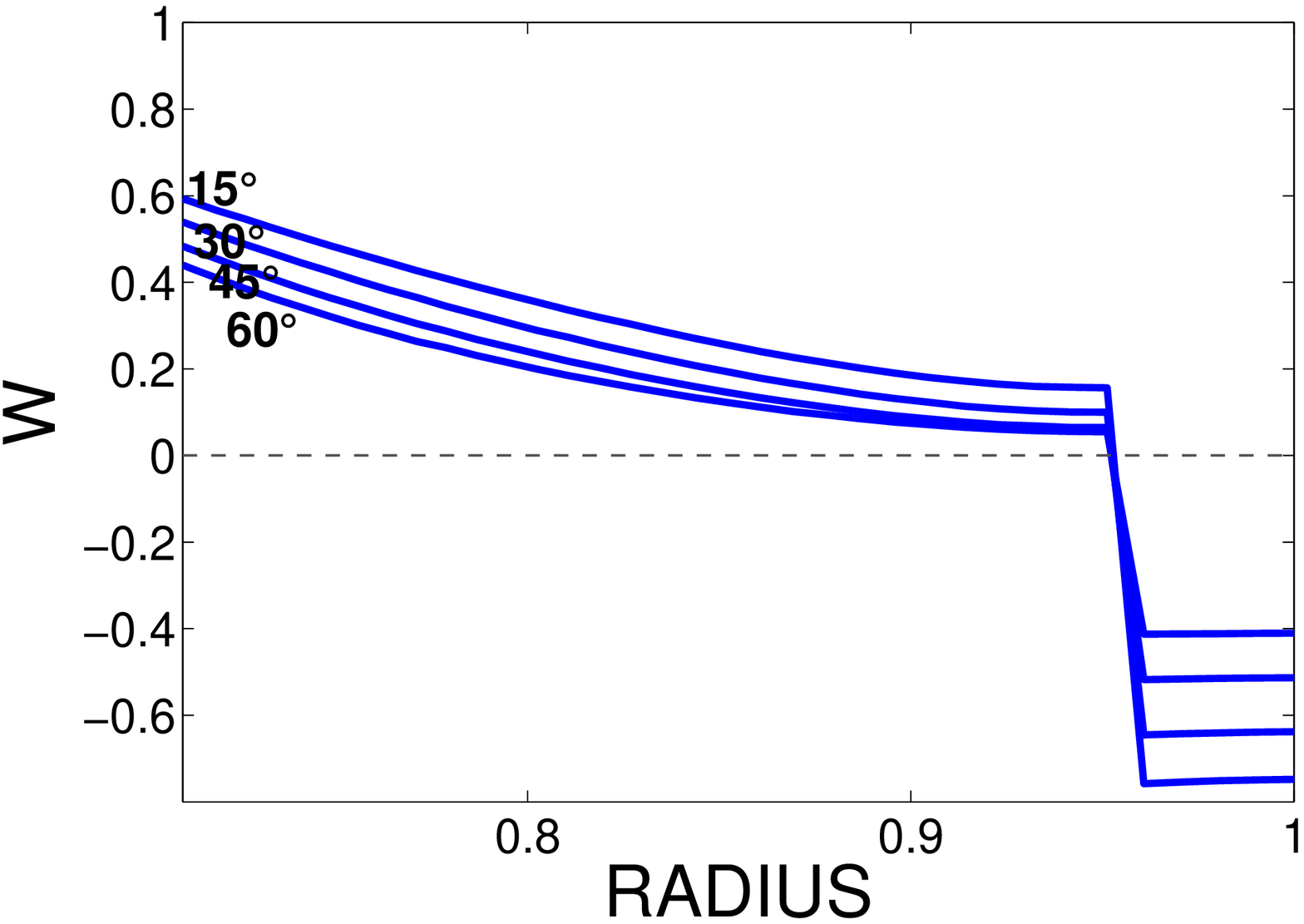}}
\end{center}
\caption[]{The same as in Fig. \ref{fig2} but  with a  density profile included which decreases outwards by more than  six scale-heights ($G=20$).}
\label{fig3}
\end{figure}

Let us solve the  mean-field equation (\ref{Rey})  in the northern hemisphere in the domain $0.6\leq x\leq 1$ with the natural boundary conditions $\partial \Om/\partial \theta=0$  at the polar and the equatorial  axis and  the stress-free condition $Q_{r\phi}=0$ at the surface. In order to model a fast tachocline-like  transition at the bottom of the domain rigid-body rotation  is there prescribed. The first-order term $V$ is assumed to exist only in the outermost layers while the third-order   effect $H$  exists down to $x=x_{\rm in}=0.7$. This simple but nonlinear model which only  ignores the transport of angular momentum by the meridional flow yields the numerical results  given in the Figs. \ref{fig1}-- \ref{fig3} which demonstrate the characteristic behavior of the horizontal stress function $W$ as the solution of the Reynolds equation (\ref{Rey}).

 We start with the homogeneous turbulence effects $V=0$ and $H=1$ in the entire convection zone. It is easy to see from (\ref{Qr}) that the outer boundary condition requires $\partial \Om/\partial r = 0$ at the equator and negative radial shear  in higher latitudes. The immediate consequence is a uniform rotation beneath the equator and a very strong subrotation along the polar axis. The equator-pole difference of $\Om$ is  always positive and it grows outwards (Fig. \ref{fig1}). It is so strong at the surface that the $W$ becomes negative there while it takes large positive values in the bulk of the convection zone.  Note, that the amplitudes of the normalized differential rotation at the surface and the $W$  at the bottom of the convection zone are numerically rather similar. Even the analytical estimates (\ref{old}) are approximately fulfilled although they are obtained with   a highly simplified linear theory.  

The rotation law in Fig. \ref{fig1}, however,    fails to deliver the characteristic superrotation of the solar equator at the bottom of the convection zone which has been helioseismologically derived  and which is the source of the `dynamo dilemma' (as it prevents the reproduction of the butterfly  diagram of the  sunspot locations within the solar cycle). It is also known, however,  that the simple model used for  Fig. \ref{fig1} does  not correctly describe the radial profiles of the turbulent $\Lambda$ terms. The linear-in-$\Om$ term $V$ mainly exists in the outer part of the convection zone while the $\Om^3$ term $H$ only exists in the inner part of the convection zone (see \R\ et al. 2013). This has consequences for the outer boundary condition which now tends to produce negative gradients of $\Om$ at all latitudes. The result for a model with  $V=-0.1$ (for $x>0.95$) and with $H=1$  for $x_{\rm in}\leq x\leq 0.95$ is  given in Fig. \ref{fig2}. Indeed, a weak equatorial superrotation is now indicated at the bottom of the convection zone. The latitudinal shear, however, is reduced so that the $W$ after (\ref{Ward})  becomes larger than in Fig. \ref{fig1}. It is again  negative in the surface region where the function $H$ is small or even zero. 

The equatorial acceleration is reduced in Fig. \ref{fig2} in comparison to the results of the `uniform' model of  Fig. \ref{fig1}. This is due to the absence of the function $H$ in the outer layers. If the weight of these layers is reduced by a strong negative  density gradient then the surface rotation law recovers and the observed value of the equator-pole difference of $\Om$ appears again (Fig. \ref{fig3}). The used density profile is simply  written as $\rho\propto \exp(-G(x-x_{\rm in}))$ which has been applied in the Reynolds equation (\ref{Rey}) with $G=20$.
\smallskip

All the calculations  lead to the result that
\begin{itemize}
\item  the equatorial acceleration is of the observed amount, 
\item  $W<0$ at the surface and  $W>0$ in the bulk of the convection zone,
\item   $W$ is  positive if averaged over the radius. 
\end{itemize}

 \section{Two  solar models}\label{Section3} 
\begin{figure}[htb]
\begin{center}
\includegraphics[width=6cm]{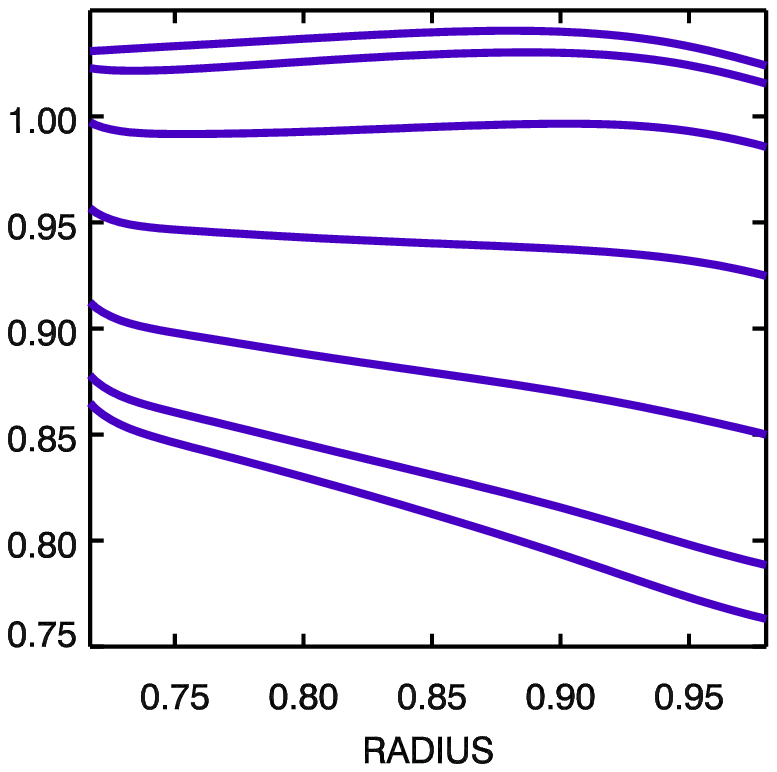}
\includegraphics[width=6cm]{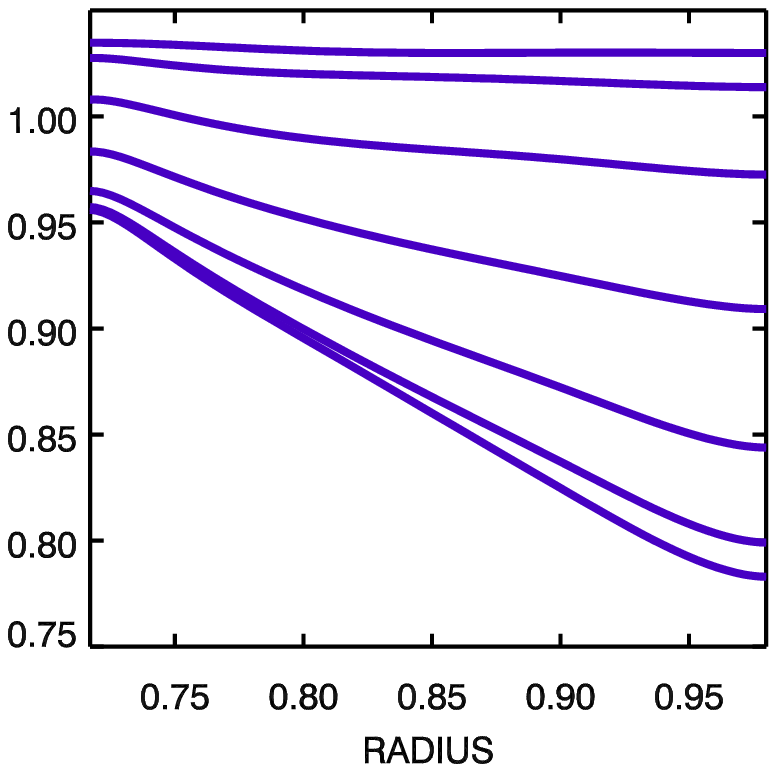}
\end{center}
\caption{The internal differential  rotation for  the full model with $\Lambda$ effect included (top) and the rotation law driven only by baroclinic flows (bottom). The meridional flow at the surface goes poleward (top) and goes equatorward (bottom). The tachocline in the theoretical models is not modeled.}
\label{rotlaw}
\end{figure}

We have also computed two models for the solar differential rotation with the detailed  physics described  by K\"uker et al.~(2011). One model includes  the full Reynolds stress tensor  while the  other one only works with (\ref{Qphi})  without $\Lambda$ (see Balbus et al. 2012). Both models are able to reproduce the surface rotation law (Fig. \ref{rotlaw}).  A convection zone model computed with the MESA stellar evolution code (Paxton et al. 2011) has been  used as a background model.  
The Reynolds stresses were computed with an average rotation period of 27 days and a value of 5/3 for the mixing length parameter. This choice sets the mixing length equal to the density scale height.    The viscosity coefficient $\nu_{\rm T}$ of  the  models is of order $10^{13}$ cm$^2$/s with a maximum 
of  $2\cdot 10^{13}$ cm$^2$/s in a depth of $x\simeq 0.8$ .   

The model without $\Lambda$ effect results in an internal rotation pattern with disk-shaped iso-contours (at the equator region). The meridional flow is directed towards the equator at the surface and towards the poles at the bottom of the convection zone (`clockwise'). The model which bases on the $\Lambda$ effect provides a meridional circulation cell with the opposite flow direction (`counterclockwise').

\begin{figure}[tbh]
\begin{center}
\includegraphics[width=7cm]{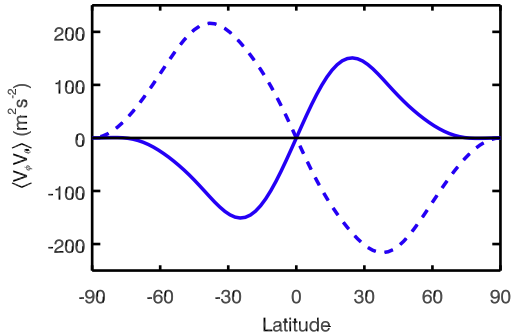}
\end{center}
\caption{The  Reynolds stress $Q_{\theta \phi}$  from the simulations and averaged over the radius  with $\Lambda$ effect (solid) and without $\Lambda$ effect (dashed). The Reynolds stress vanishes at the equator and changes the sign in the southern hemisphere (see Fig. \ref{fig0}).}
\label{full_reynolds}
\end{figure}

Figure \ref{full_reynolds}  gives the horizontal Reynolds stress $Q_{\theta \phi}$ averaged over the radius in its dependence on the  latitude. This quantity is antisymmetric  with respect to the equator by definition and thus vanishes at the equator. If the Reynolds stress is purely viscous (dashed line)  $Q_{\theta \phi}$ is negative (positive)   in the northern (southern) hemisphere -- in contrast to the empirical findings  given in Fig. \ref{fig0}. For the model with the full Reynolds stress (solid line)  the horizontal $\Lambda$ effect exceeds the viscous part in the bulk of the convection zone where the total Reynolds stress is positive (negative) in the northern (southern) hemisphere.  As it must, the horizontal stress $Q_{\theta \phi}$ vanishes both on the rotation axis and in the equatorial plane. It reaches its peak value at about $30^\circ$ latitude.  The amplitude of the horizontal cross correlation $Q_{\theta \phi}$ is  of order $10^6$ cm$^2$/s$^2$ which leads to $|W| \simeq 0.14$  well corresponding to the estimate  (\ref{old})$_2$  as from the quasilinear turbulence theory $H\simeq 0.4$ results. The value  in very good  agreement with  the numerical results for the cubic Reynolds equation given in Fig.  \ref{fig3}.

We have  to note that the numerical value of the positive maximum of $10^6$ cm$^2$/s$^2$ exceeds the empirical results by a factor of five which might be a consequence of a too high viscosity value used in the simulations.
\section{Conclusions}
It is hard to imagine  that the mechanism of the  solar dynamo could  be understood without understanding of the maintenance of the solar rotation law. It is thus of  relevance if one can show that the generation of the differential  rotation of the solar convection zone bases on the existence of nondiffusive $\Lambda$ terms in the Reynolds stress in a similar sense as the solar dynamo may base  on the existence of nondiffusive $\alpha$  terms in the turbulent electromotive force.
We argue that indeed the $\Lambda$ effect in rotating anisotropic turbulence  is {\em necessary} to explain the current observation of positive (negative) horizontal cross correlation  $Q_{\theta \phi}$ at the northern (southern) hemisphere  at the solar surface which Hathaway et al.  indicated  as a result of the proper motions of giant cells.  All analytical and numerical studies lead to positive functions $H$  which in the bulk of the convection zone are able to overcompensate the diffusive term which alone would lead to the opposite signs which are not observed.  The overcompensation, however, is not trivial. It might be  that in the outer supergranulation layer ($0.95\lsim x\lsim 1$)  the equatorial acceleration is too large and/or the $H$ is too small. We have to mention in this respect that the $H$ is a term which is of higher order in the Coriolis number $\Om^*= 2 \tau_{\rm corr}\Om$ so that due to the short lifetimes of the granulation and supergranulation the $H\ll 1$ in the outer part of the convection zone and the turbulent medium behaves diffusive in the horizontal plane. Because of symmetry properties a term linear in $\Om$ does not exist in the horizontal cross-correlation $Q_{\theta\phi}$ -- in opposition to the radial tensor component $Q_{r\phi}$. 

We have  demonstrated  also by means of a simplified model which ignores  the meridional flow in the solution of the azimuthal Reynolds equation   that $\cos\theta Q_{\theta \phi}$ --  or which is the same -- the quantity $W$  is indeed negative in the surface layers.  The negative sign, however, only describes a surface effect. In subsurface layers the equator-pole difference of $\Om$ is reduced and the amplitude of the positive $H$ grows. It is thus no surprise that in the deeper layers of the  convection zone the  sign of $W$ becomes  positive. This is a general result which does not depend on details of the eddy viscosity, of the density stratification, and/or influences of the meridional flow. Even our most complex $\Lambda$ effect model which will  reproduce the internal solar  rotation law and also the observed pattern of the meridional flow exactly shows the described behavior. If the equatorial acceleration is {\em not} a consequence of the $\Lambda$ effect but  is due to a meridional flow (which  {\em must}  flow equatorward along the surface) then the function $W$ would be  negative-definite on the northern hemisphere through the entire convection zone .

%
%

\begin{thebibliography}{}
\bibitem[\protect\citeauthoryear{}{}]{}
Balbus, S.H., Latter, H. \& Weiss, N. 2012, MNRAS, 420, 2457
\bibitem[\protect\citeauthoryear{}{}]{}
Balthasar,  H., Vazquez, M., \& W\''ohl, H 1986, A\&A, 155, 87
\bibitem[\protect\citeauthoryear{}{}]{}
Boussinesq, M.J. 1887, Th´eorie de l’´ecoulement tourbillonnant et tumultueux des liquides, Gauthier-Villars et fils, Paris 
\bibitem[\protect\citeauthoryear{}{}]{}
Chan, K.L. 2001, ApJ, 548, 1102
\bibitem[\protect\citeauthoryear{}{}]{}
Gilman, P.A., \& Howard, R.   1984, Solar. Phys. 93, 171
\bibitem[\protect\citeauthoryear{}{}]{}
Gizon, L., \&  Rempel, M. 2008,  Sol. Phys  251, 241
\bibitem[\protect\citeauthoryear{}{}]{}
 Hathaway, D.H., Upton, L., \&  Colegrove, O. 2013, Science, 342, 1217
\bibitem[\protect\citeauthoryear{}{}]{}
K\"appyl\"a, P.J., Mantere, M.J., Guerrero, G., et al. 2011, A\&A 531, A162
\bibitem[\protect\citeauthoryear{}{}]{}
Kippenhahn, R. 1963, ApJ, 137, 664
\bibitem[\protect\citeauthoryear{}{}]{}
K\"uker, M., R\"udiger, G., \& Kitchatinov, L.L. 2013 A\&A, 530, 48
\bibitem[\protect\citeauthoryear{}{}]{}
Nesme-Ribes, E., Ferreira, E.N. \& Vince, L.  1993, A\&A, 276, 211
\bibitem[\protect\citeauthoryear{}{}]{}
Paxton, B., Bildsten, L., Dotter, A., et al. 2011, ApJS, 192, 3
\bibitem[\protect\citeauthoryear{}{}]{}
Reinhold, T., Reiners, A., \& Basri, G. 2013, A\&A, 560, 4
\bibitem[\protect\citeauthoryear{}{}]{}
\R, G. 1989, Differential rotation and solar convection, Gordon \& Breach Science Publishers
\bibitem[\protect\citeauthoryear{}{}]{}
\R, G., Kitchatinov, L.L.,  \& Hollerbach, R. 2013, Magnetic processes in astrophysics: theory, simulations, experiments. Wiley-VCH Weinheim
\bibitem[\protect\citeauthoryear{}{}]{}
Schad, A., Timmer, J., \& Roth, M. 2012, Astron. Nachr. 333, 991
\bibitem[\protect\citeauthoryear{}{}]{}
Taylor, G.I. 1915, Phil. Trans. Roy. Soc.,  A215, 1
\bibitem[\protect\citeauthoryear{}{}]{}
Vrsnak, B., Brajsa, R., W\''ohl, H. et al. 2003, A\&A, 404, 1117
\bibitem[\protect\citeauthoryear{}{}]{}
Ward, F. 1965, ApJ, 141, 534

 \end{thebibliography}
%

\end{document}